\documentclass{webofc}
\usepackage[varg]{txfonts}   
\usepackage{bbold}

\newcommand{\intq}{\int dp_4\,}
\newcommand{\pslash}{p\hspace{-1.8mm}\slash}
\newcommand{\vecpslash}{\vec p\hspace{-1.8mm}\slash}

\begin{document}
\title{Gauge-fixed Lattice QCD and the dispersion relation of Wilson fermions}
%
%

\author{\firstname{Giuseppe} \lastname{Burgio}\inst{1}\fnsep\thanks{\email{giuseppe.burgio@uni-tuebingen.de}} \and
        \firstname{Hannes} \lastname{Vogt}\inst{1}\fnsep\thanks{\email{hannes@vogt.de}} 
}

\institute{Institut für Theoretische Physik, Auf der Morgenstelle 14, 72076 Tübingen (Germany)
          }

\abstract{%
  We show that, when investigating Wilson-fermions correlation functions on the lattice, 
  one is bound to encounter major difficulties in defining their 
  dispersion relation, even at tree level. The problem is indeed quite general and, although we stumbled upon it while
  studying Coulomb-gauge applications, it also affects gauge fixed studies in covariant gauges, including 
  their most popular version, Landau gauge.
  In this paper we will discuss a solution to this problems based on a redefinition of the 
  kinematic momentum of the fermion.
}
\maketitle
\section{Introduction}
\label{intro}
The investigation of the IR-properties of non abelian gauge theories is still one of the
most interesting and fruitful chapters of particle physics. In particular, lattice studies have
contributed a great deal to our understanding of the gauge invariant aspects of confinement,
chiral symmetry breaking, hadron spectrum etc. However, both from a theoretical and from 
a practical point of view, there exists a broad interest in analysing whether and how such properties are  
encoded in gauge-fixed Green's functions, mainly since most applicable continuum methods still heavily rely on them.
The corresponding first studies for QCD 
date back to the '70s, see e.g. 
\cite{Casher:1974vf,Gribov:1977wm,Mandelstam:1979xd,Kugo:1979gm,Cornwall:1981zr}, and
many analytic and numerical investigations have since followed, both in covariant 
\cite{Zwanziger:1991gz,Zwanziger:1992qr,Zwanziger:1993dh,Roberts:1994dr,Cucchieri:1995pn,Cucchieri:1997dx,%
Cucchieri:1997ns,Leinweber:1998uu,Skullerud:2000un,Alkofer:2000wg,Boucaud:2000ey,Bonnet:2001uh,Bowman:2001xh,Bonnet:2002ih,%
Bowman:2002bm,Zwanziger:2003cf,Bloch:2003sk,Bowman:2004jm,Gattnar:2004bf,Cucchieri:2004sq,Bogolubsky:2005wf,Schleifenbaum:2006bq,%
Bogolubsky:2007bw,Dudal:2009xh,%
Vandersickel:2012tz,Schaden:2013ffa}
and non-covariant gauges 
\cite{Dell'Antonio:1991xt,Zwanziger:1998ez,Cucchieri:2000gu,Cucchieri:2000hv,Cucchieri:2000kw,Zwanziger:2002sh,%
Feuchter:2004mk,Greensite:2004ke,Epple:2006hv,Nakagawa:2006fk,Quandt:2007qd,Epple:2007ut,Quandt:2008zj,Reinhardt:2008ek,Burgio:2008jr,%
Burgio:2009xp,Campagnari:2010wc,Leder:2010ji,Quandt:2010yq,%
Reinhardt:2011fq,Reinhardt:2011hq,Heffner:2012sx,Burgio:2012bk,Burgio:2012ph,Pak:2013uba,Vogt:2013jha,Greensite:2014bua,%
Burgio:2015hsa,Campagnari:2015zsa,Heffner:2015zna,Vastag:2015qjd,Burgio:2016nad,Reinhardt:2017pyr}. 
A consistent picture emerges from this corpus, supporting Gribov's original idea \cite{Gribov:1977wm}: 
restricting the functional integral to the first Gribov region indeed generates a dynamical mass scale 
${\cal{O}}\left(1\,\mathrm{GeV}\right)$. Coulomb-gauge in particular, being the closest we have to a physical
description \cite{Christ:1980ku,Adler:1984ri}, offers many advantages; e.g., one can show
that the self energy of the gluon can in fact be described with Gribov's original formula $\omega_A = \sqrt{|\vec{p}|^2 + M^4/|\vec{p}|^2}$,
again with $M = {\cal{O}}\left(1\,\mathrm{GeV}\right)$ \cite{Burgio:2008jr}. 

A lesson one constantly learns from all lattice investigations at fixed gauge is that, to be able to extract meaningful informations,
one must take particular care in treating discretization artifacts, which often cloud the sought results.
For example, the gluon self-energy above could only be obtained after devising a proper scheme
to remove spurious contributions caused by the discretized time \cite{Burgio:2008jr}. Similarly, extracting the 
Coulomb string tension directly from the Coulomb-kernel, which one naively would consider the most
straightforward prescription, turns out to be a quite non trivial task; alternative, less intuitive definitions
turn out to be more efficient \cite{Burgio:2015hsa}. This holds of course also for quark-quark correlation functions,
as the extraction of the chiral mass and of the self energy for staggered fermions, 
necessary to detect any evidence of chiral symmetry breaking and confinement, 
clearly exemplifies \cite{Burgio:2012ph}.

In this paper we will show that for Wilson fermions extra care needs to be taken when 
calculating fermionic Green's functions. Indeed, the problem lies in the very definition of a suitable kinematic momentum 
against which to plot fermionic correlation functions, independently of the gauge, and thus affects 
all gauge fixed investigation, e.g. also in Landau gauge \cite{Skullerud:2000un}.
Of course, one could argue that other lattice fermions could be used for QCD instead, since they should
not suffer from such problems. However, the recent surge in investigations of beyond the standard model scenarios,
with many lattice simulations performed for different models, has increased the interest in
studying the non perturbative properties of correlation functions in QCD-like theories; depending on the model chosen, 
one will want to keep the freedom of using the most convenient discretization for the fermionic fields.
Indeed, that's how we noticed the issue in the first place - when trying to apply the analysis of \cite{Burgio:2012ph}
to SU(2) + (adjoint) Wilson fermions configurations generated by the Edinburgh-Swansea-Odense collaboration 
as a model for minimal walking Technicolor \cite{DelDebbio:2008zf,DelDebbio:2010hu,DelDebbio:2010hx,DelDebbio:2015byq}.


\section{The fermion self energy}
\label{sec-1}

The most general structure for the expectation value of the Dirac operator, 
both in covariant and Coulomb gauges, reads \cite{Skullerud:2000un,Burgio:2012ph}:
\begin{equation}
S^{-1}(\vec{p},p_4) = i\, \vecpslash A_s(\vec{p},p_4) +i\, \pslash_4 A_t(\vec{p},p_4) + i\, p_i \gamma_i p_4 \gamma_4 
A_d(\vec{p},p_4) + B(\vec{p},p_4) \mathbb{1}\,.\label{eq-1}
\end{equation}
$A_d$ can be shown to vanish in general, while of course 
$A_s (p)= A_t (p)$ for covariant gauges \cite{Skullerud:2000un,Burgio:2012ph}. For Coulomb gauge, moreover, $A_s$, $A_t$ and $B$ can be
shown to be $p_4$ independent \cite{Burgio:2012ph}. Factorizing $A_s =Z^{-1}$ out, Eq.~(\ref{eq-1}) can be rewritten as \cite{Skullerud:2000un,Burgio:2012ph}: 
\begin{equation}
S^{-1}(\vec{p},p_4) =Z^{-1}(\vec{p}) \left[ i\, \vecpslash 
+ i\, \pslash_4 \, \alpha(\vec{p})+M(\vec{p}) \mathbb{1} \right] \quad
\left(\mathrm{cov.}\;S^{-1}(p) =Z^{-1}({p}) \left[ i\, \pslash +M({p}) \mathbb{1} \right]\right)\,,\label{eq-2}
\end{equation}
where $\alpha = A_t/A_s$ and $M = B/A_{s}$. The inverse Dirac propagator above can be used to define the static quark
propagator $S(\vec{p}) = \intq S(\vec{p},p_4)$, which, in analogy to the free case, corresponds up to a factor 
$1/2$ to the Hamiltonian operator $H$. 
Contrary to covariant gauges, this is straightforward to calculate in Coulomb gauge, yielding \cite{Burgio:2012ph}:
\begin{equation}
H(\vec{p}) = \frac{\displaystyle Z(\vec{p})}{\displaystyle 
        \alpha(\displaystyle \vec{p})} \frac{\displaystyle \sqrt{\vec{p}^2 
          + M^2(|\vec{p}|)}}{\displaystyle i\, \vecpslash + M(\vec{p}) \mathbb{1}}= 
      \frac{\displaystyle Z(\vec{p})}{\displaystyle 
        \alpha(\displaystyle \vec{p})} \frac{\displaystyle -i\, \vecpslash +
          M(\vec{p}) \mathbb{1}}{\displaystyle \sqrt{\vec{p}^2 + M^2(|\vec{p}|)}}\,.\label{eq-3}
\end{equation}
The dispersion relation of the fermion can therefore be directly read off from Eq.~(\ref{eq-3}), since the inverse
coefficient of $-i\, \vecpslash + M(\vec{p)}$ corresponds to the eigenvalue of $H$:
\begin{equation}
\omega (|\vec{p}|)= \frac{\alpha(|\vec{p}|)}{Z(|\vec{p}|)}
      \sqrt{\vec{p}^2 + M^2(|\vec{p}|)}\,.\label{eq-4}
\end{equation}
As shown in \cite{Burgio:2012ph}, chiral symmetry breaking is encoded in $M(|\vec{p}|)$, while for a confining theory both 
$\omega_A$ and $\omega$ should be IR-divergent. Non-perturbative Coulomb
gauge therefore nicely decouples the chiral symmetry and confinement properties of QCD-like theories,
making it an interesting tool in the analysis of beyond the standard model scenarios \cite{Cossu:2008wh}.

\section{The kinematic momentum for Wilson fermions}
\label{sec-2}
If the self-energy in Eq.~(\ref{eq-4}) had been obtained in the continuum, there would be nothing
one should be careful about. On the lattice, on the other hand, the discrete momenta $k_i = \frac{2\pi}{aN_i}n_i$,
$k_4=\frac{2\pi}{aN_t}\left(n_4+\frac{1}{2}\right)$ do not define suitable kinematic variables that
can be related to the continuum ones. For staggered and overlap fermions
there is however an easy prescription: calculate first, either analytically or numerically, the quark propagator 
in the free case, 
$S^\text{(0)} (k) = i\, \left( \sum_\mu C_\mu(k_\mu) \gamma_\mu \right) + m \,\mathbb{1}$. 
Isolating the coefficients of
the (euclidean) $\gamma$-matrices defines now the kinematic momenta: $p_\mu(k_\mu) = C_\mu(k_\mu)$. 
The reason why such procedure works is of course that the $\mathbb{1}$-coefficient is 
$k$ independent. For Wilson fermions, on the other hand, the doubling problem is
solved by modifying exactly such coefficient, introducing an explicit $k$-dependence:
\begin{equation}
S^\text{(0)} (k) = \frac{i\,}{a}\sum_{\mu=1}^4\gamma_\mu \tilde p_\mu (k) + \left(m+ \frac{a}{4}\sum_{\mu=1}^4 p_\mu^2 (k)\right)\,\mathbb{1}\,,\label{eq-5}
\end{equation}
with $\tilde{p}_\mu (k_\mu)= \frac{1}{a} \sin(a k_\mu)$ and $p_\mu(k_\mu) = \frac{2}{a} \sin(a k_\mu/2)$; a definition of the 
kinematic momenta as for staggered/overlap fermions is therefore not possible. Previous studies in Landau gauge \cite{Skullerud:2000un} have 
ignored the problem and used
$p_\mu$ instead; this however does not reproduce the correct dispersion relation even in the free case, in contrast e.g.
to $\omega_A$, which is a "natural" function of $p_\mu(k_\mu)$ \cite{Leinweber:1998uu}. 

\begin{figure}[htb]
\centering
\includegraphics[width=0.8\linewidth]{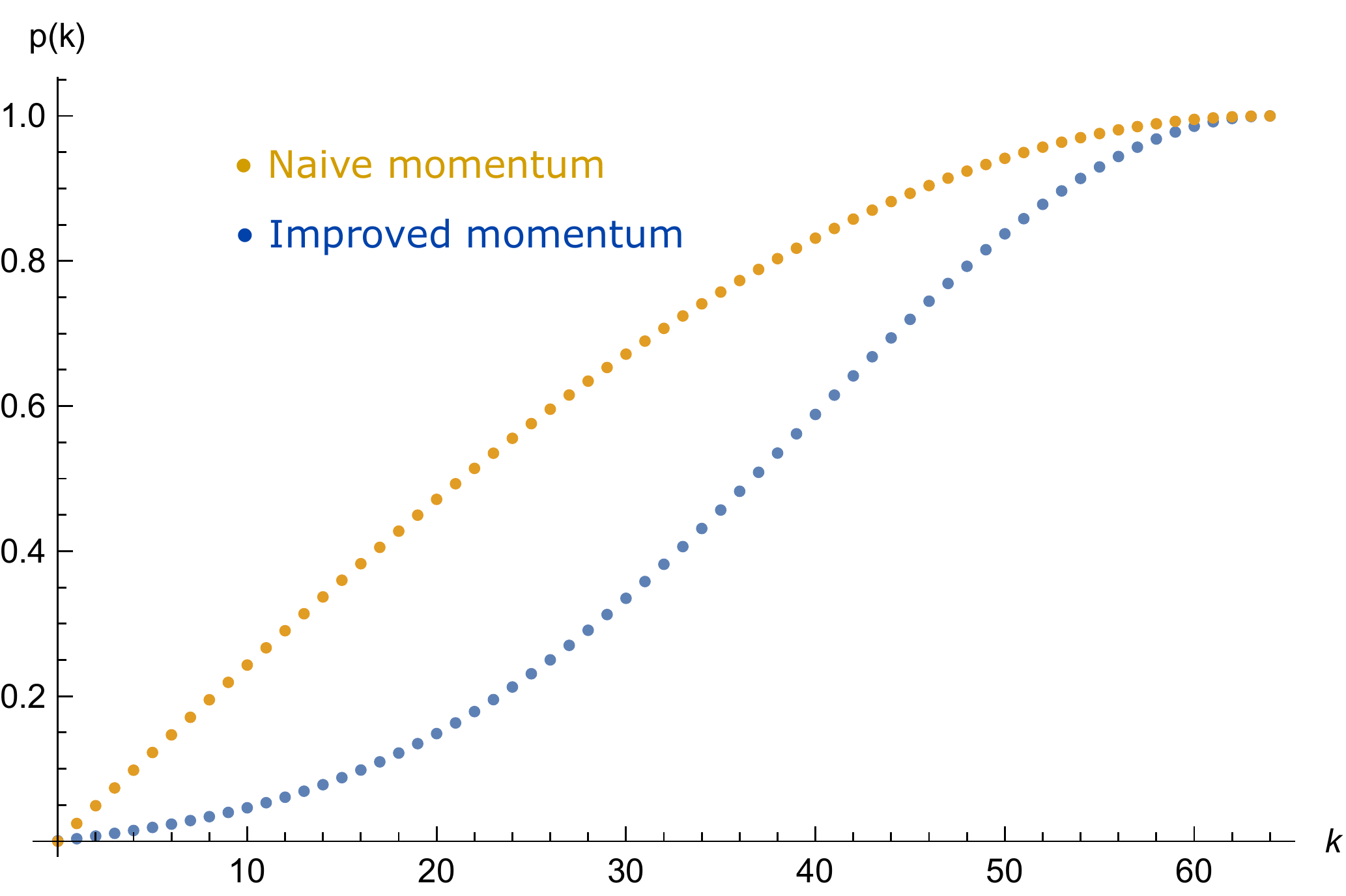}
\caption{Comparison between $p(k) = 2 \sqrt{\sum_i\sin^2(k_i/2)}$ and Eq.~(\ref{eq-14}) for a $128^4$ lattice ($\vec{k} = (k,k,k)$).}
\label{fig-1}       
\end{figure}

\begin{figure}[htb]
\centering
\includegraphics[width=0.8\linewidth]{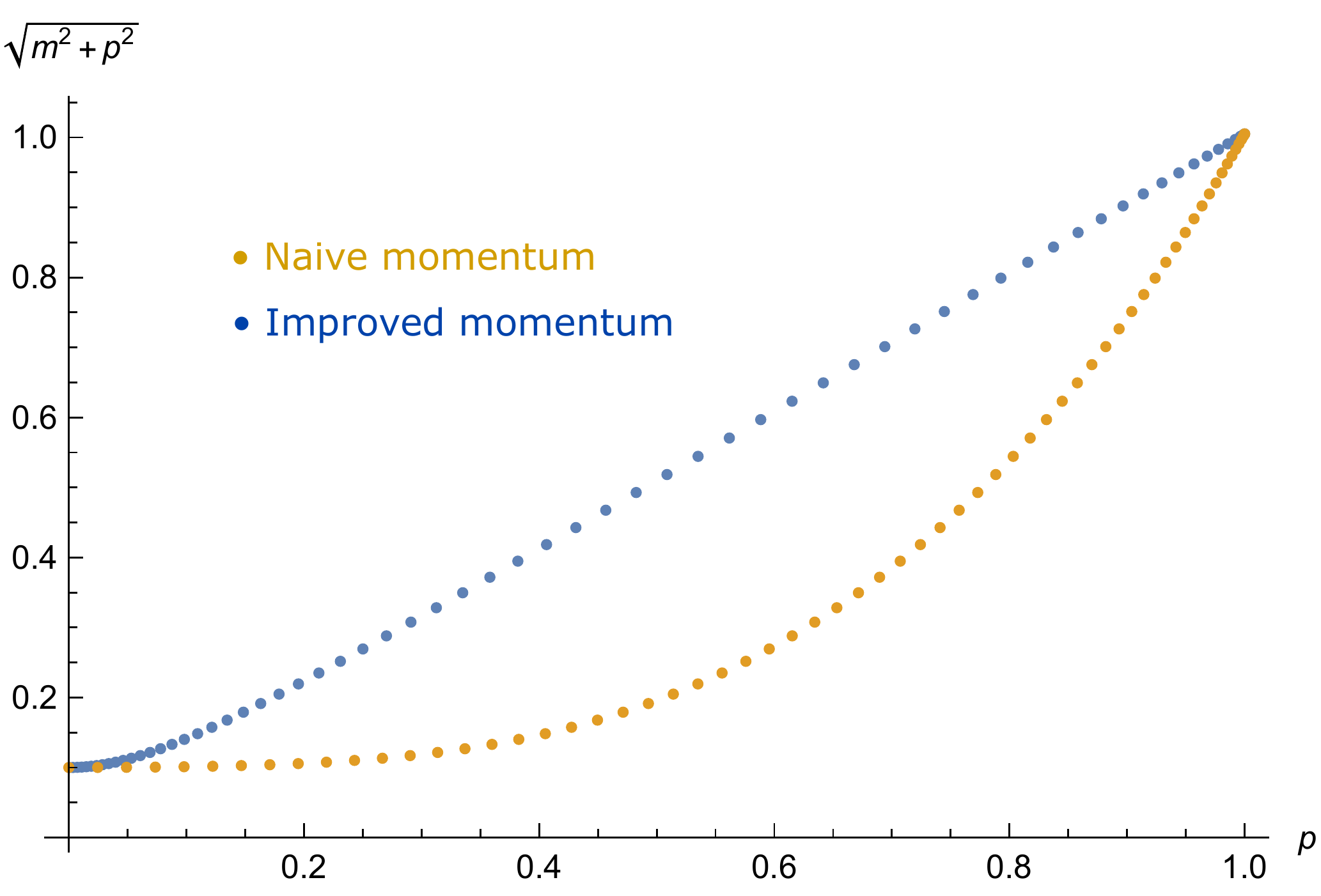}
\caption{Massive fermionic dispersion relations vs.
naive and improved momentum from Eq.~(\ref{eq-14}).}
\label{fig-2}       
\end{figure}

To illustrate our proposal, start by considering the free static propagator:
\begin{equation}
 \int_{-\infty}^\infty \frac{dp_4}{2\pi} S_0(\vec{p},p_4) = \frac{-i\, 
{\vecpslash}+m\,\mathbb{1}}{2\sqrt{\vec{p}^2+m^2}}\equiv \frac{-i\, {\vecpslash}+m\,\mathbb{1}}{2\omega_{0}}\,,\label{eq-6}
\end{equation}
where $\omega_{0}$ is the free dispersion relation, i.e. the eigenvalue of the free Hamiltonian,
the $p$ are all functions of $k$ and:
\begin{equation}
S_0(\vec{p},p_4) = \frac{-i\, \pslash_4 -i\, {\vecpslash}+m\,\mathbb{1}}{p_4^2+\vec{p}^2+m^2}\,.\label{eq-7}
\end{equation}
Notice now that the free dispersion relation can be directly obtained from the free propagator:
\begin{equation}
\omega_{0}^{-1} (k) = \frac{1}{\sqrt{\vec{p}^2(k)+m^2}}= \int_{-\infty}^\infty \frac{dp_4}{2\pi} \left|S_0(\vec{p}(k),p_4)\right|^2\,,\label{eq-8}
\end{equation}
where
\begin{equation}
\left|S_0(p)\right|^2 = S_0^\dag(p) S_0(p) = \frac{1}{p_4^2+\vec{p}^2+m^2}\,,\label{eq-9}
\end{equation}
The kinematic momentum can be thus "defined" via $S_0$, calculating $\omega_{0} (k)$ 
through Eq.~(\ref{eq-8}) and setting $|\vec{p}(k)| = \sqrt{\omega_{0}^2(k)-m^2}$, very much in
the spirit of the gluonic "tree-level correction" of \cite{Leinweber:1998uu}.
Since in lattice calculations, to avoid discretization errors arising from the breaking of
rotational invariance, one only considers diagonal momenta of the form $\vec{k} = (k,k,k)$ \cite{Leinweber:1998uu}, 
such prescription is sufficient for all practical purposes.
Explicitly (for the sake of readability the lattice spacing will be set to $a=1$ in the following):
\begin{align}
S_0^\text{lat}(k) &= \frac{-i\, \gamma_4 \sin(k_4) -i\, \sum_i \gamma_i \sin(k_i) + (m + 
\sum_\mu (1-\cos(k_\mu)))\,\mathbb{1}}{\sin^2(k_4)+\sum_i \sin^2(k_i)+\left[m+\sum_\mu (1-\cos(k_\mu))\right]^2}\,,\label{eq-10}\\
\left|S_0^\text{lat}(k)\right|^2 & = \frac{1}{\sin^2(k_4)+\sum_i \sin^2(k_i)+\left[m+\sum_\mu (1-\cos(k_\mu))\right]^2}\,.\label{eq-11}
\end{align}
Replacing ${\displaystyle{\int \frac{dp_4}{2\pi}}}$ with ${\displaystyle{\frac{1}{2\pi}\frac{4\pi}{N_t}\sum_{k_4}}}$ in Eq.~(\ref{eq-8}) we thus get:
\begin{align}
\omega_{0,m} ^{-1}(k) & = \frac{1}{\sqrt{\vec{p}^2+m^2}} \equiv \frac{2}{N_t}\sum_{k_4} \left|S_0^\text{lat}(k)\right|^2\,,\label{eq-12}\\
|\vec{p}({k})| & = \sqrt{\frac{1}{\left(\frac{2}{N_t}\sum_{k_4} \left|S_0^\text{lat}(k)\right|^2\right)^2}-m^2}\,.\label{eq-13}\
\end{align}
Setting now $m=0$ we get, as operative definition of the kinematic momentum:
\begin{align}
  |\vec{p}(k)|^{-1} &=\frac{2}{N_t} \sum_{k_4} \left|S_0^\text{lat}(k)\right|^2\nonumber\\
  &= \frac{2}{N_t}  \sum_{n_4 = -N_t/2+1}^{N_t/2} \frac{1}{\sin^2(k_4)+\sum_i \sin^2(k_i)+\left[\sum_\mu 
(1-\cos(k_\mu))\right]^2}\,.\label{eq-14}
\end{align}
The values arising from Eq.~(\ref{eq-14})
can be calculated once and for all for each desired lattice size and stored to be used whenever
needed. Figure~\ref{fig-1} shows the comparison between the "naive" fermionic
kinematic momentum $p(k) = 2 \sqrt{\sum_i\sin^2(k_i/2)}$, as used e.g. in \cite{Skullerud:2000un},
and our definition Eq.~(\ref{eq-14}) for a $128^4$ lattice. As one can see, the differences are quite striking throughout
the whole Brillouin-zone.
Figure~\ref{fig-2} compares the (massive) free fermionic dispersion relations plotted against
the naive momentum and the one obtained from Eq.~(\ref{eq-14}), with an arbitrarily chosen bare mass. While the former shows
a dependence which is physically meaningless for all momenta and is bound to distort any IR results, 
the latter gives (by definition!) exactly the
expected physical behaviour throughout the whole Brillouin-zone.

The above definition will suffice for most applications.
Should one however think that mass contributions in the definition 
of the kinematic momenta might play a role in the application at hand, and 
therefore wish to extend the prescription to
$m\neq0$, a bit of care is needed. 
As one can immediately verify, for $m\neq0$ Eq.~(\ref{eq-12}) shifts
the pole from $m$ to $\frac{1}{2}m^2+m$, 
so that from Eq.~(\ref{eq-13}) one gets $\vec p(k = 0) \neq 0$, 
moreover with the wrong slope at the origin, $\left.\displaystyle{\frac{{d}{p}}{{d}{k}}}\right|_{k=0} \neq 1$.
A "mass-improved" prescription should be therefore defined as follows.
Expand first $\omega^2_{0,m}(k)$ from Eq.~(\ref{eq-12}) around $k=0$ up to $2^{\mathrm{nd}}$ order for diagonal 
momenta, obtaining:
\begin{equation}
\omega_{0,m}^2(k) = \left[ \frac{N_t}{2} \frac{1}{\sum_{k_4} 
\frac{1}{\sin^2\left(k_4\right)+3\sin^2\left(k\right)+\left[m+2\sin^2\left(\frac{k_4}{2}\right)+6\sin^2\left(\frac{k}{2}
\right)\right]^2}} \right]^2
= \omega^2_{0,m}(0) + \frac{1}{2} \left. \frac{\partial^2 \omega_{0,m}^2(k)}{\partial k^2} \right|_{k=0} k^2+ \mathcal O(k^4)\,.\label{eq-15}
\end{equation}
From this on can then directly define:
\begin{equation}
|\vec{p}(k)|^2 = \left[\omega_{0,m}^2(k) - \omega_{0,m}^2(0) \right]\left[ \frac{1}{2} \left. \frac{\partial^2 
\omega_{0,m}^2(k)}{\partial 
k^2} \right|_{k=0}\right]^{-1}\,,
\end{equation}
with:
\begin{align}
\omega^2_{0,m}(0) &= \left[ \frac{N_t}{2 \sum_{k_4} \frac{1}{S_m(k_4)}}\right]^2\,,  \quad\qquad
\left. \frac{\partial^2 \omega_{0,m}^2(k)}{\partial k^2} \right|_{k=0}
 = \frac{N_t^2 \sum_{k_4} 
\frac{6+6\left(m+2\sin^2\left(\frac{k_4}{2}\right)\right)}{S^2_m(k_4)}}{2 \left[\sum_{k_4} \frac{1}{S_m(k_4)}\right]^3}\,,\nonumber\\
 S_m(k_4) &= \sin^2(k_4) + \left[ m+2 \sin^2\left(\frac{k_4}{2}\right) \right]^2= m^2+4(m+1)\sin^2\left(\frac{k_4}{2}\right)\,.\label{eq-17}
\end{align}

\section{Conclusions and outlook}
\label{concl}
In this paper we have given an improved definition of the kinematic momentum for
Wilson fermions, which we believe to be crucial if one
wishes to extract any information from the IR-behaviour of correlation functions. 
Such definition, adapted accordingly, should be used in any gauge-fixed analysis involving Wilson-fermions,
e.g. also in Landau gauge. 
Concerning our original motivation, i.e. the application of the analysis of \cite{Burgio:2012ph}
to the configurations used in \cite{DelDebbio:2008zf,DelDebbio:2010hu,DelDebbio:2010hx,DelDebbio:2015byq},
two further problem still need to be solved. First, for Wilson fermions the coefficient $B$ in Eq.~(\ref{eq-1}) is $k_4$ dependent, both 
at tree level and in the interacting case. Normalizing $B$ to its tree level expression, i.e. the $\mathbb{1}$
coefficient in Eq.~(\ref{eq-5}),
as proposed in \cite{August:2013jia} for Landau gauge, could only work if
the ratio for different energies at fixed $\vec k$, 
$ \displaystyle{{B_\text{lat}(\vec k,k^1_4)}/{B_\text{lat}(\vec k,k^2_4)}}$ with $k^1_4 \neq k^2_4$,
would exactly match their equivalent ratios at tree-level, which one can immediately verify not to be the case. Thus
a different prescription needs to be developed, a possible one being to average the
normalized ratios over $k_4$: 
\begin{equation}
 B(\vec k) = m \frac{1}{N_t} \sum_{k_4} \frac{B_\text{lat}(\vec k,k_4)}{B^0_\text{lat}(\vec k,k_4)}\,,\label{eq-18}
\end{equation}
where the factor $m$ is needed to guarantee $B(0) = m$. 
The second problem comes from the fact that for adjoint Wilson fermions the
massless limit at $\beta = 2.25$ is approached for a negative bare mass, $-am_0 \approx 1.20$. 
However this is neither a meaningful choice in Eq.~(\ref{eq-18})
nor does it provide a reasonable tree-level correction for $B$, as 
already pointed out for Landau gauge \cite{August:2013jia}. 
The proposal there was to use the PCAC mass instead, which might be a good choice in the continuum limit, 
but can still be the source of other uncertainty at finite lattice spacing. Further work will definitely
be needed to settle the issue.

%
%
%
%
\bibliography{thesis}
%
%
%
%

\end{document}